\documentclass[twocolumn,dvipsnames]{aastex63}
\usepackage{gensymb}
\usepackage{amsmath}
\usepackage{amssymb}
\usepackage{xcolor}

\usepackage{pifont}

\hypersetup{colorlinks=true, linkcolor=red, filecolor=magenta, urlcolor=Sepia, citecolor=blue}

\definecolor{ForestGreen}{HTML}{228B22}

\received{XXX}
\revised{YYY}
\accepted{ZZZ}
\submitjournal{ApJ}

\shorttitle{Detection of Super-Chandrasekhar white dwarfs by gravitational wave }
\shortauthors{Kalita, Mukhopadhyay, Mondal \& Bulik}

\begin{document}

\title{Timescales for detection of super-Chandrasekhar white dwarfs by gravitational wave astronomy}


\author[0000-0002-3818-6037]{Surajit Kalita}
\affiliation{Department of Physics, Indian Institute of Science, Bangalore 560012, India}
\email{surajitk@iisc.ac.in}

\author[0000-0002-3020-9513]{Banibrata Mukhopadhyay}
\email{bm@iisc.ac.in}
\affiliation{Department of Physics, Indian Institute of Science, Bangalore 560012, India}

\author[0000-0001-8174-2011]{Tushar Mondal}
\email{mtushar@iisc.ac.in}
\affiliation{Department of Physics, Indian Institute of Science, Bangalore 560012, India}

\author{Tomasz Bulik}
\email{tb@astrouw.edu.pl}
\affiliation{Astronomical Observatory, University of Warsaw, Al. Ujazdowskie 4, 00478 Warszawa, Poland}
\begin{abstract}

In about last couple of decades, the inference of the violation of the Chandrasekhar mass-limit of white dwarfs from indirect observation is probably a revolutionary discovery in astronomy. Various researchers have already proposed different theories to explain this interesting phenomenon. However, such massive white dwarfs usually possess very little luminosity, and hence they, so far, cannot be detected directly by any observations. We have already proposed that the continuous gravitational wave may be one of the probes to detect them directly, and in the future, various space-based detectors such as LISA, DECIGO, and BBO, should be able to detect many of those white dwarfs (provided they behave like pulsars). In this paper, we address various timescales related to the emission of gravitational as well as dipole radiations. This exploration sets a timescale for the detectors to observe the massive white dwarfs.
\end{abstract}

\keywords{White dwarf stars (1799) --- Pulsars (1306) ---  Gravitational waves (678) --- Astronomical radiation sources (89) --- Stellar magnetic fields (1610) --- Chandrasekhar limit (221)}

\section{Introduction}\label{Introduction}

In white dwarfs (WDs), the inward pressure due to gravity balances the outward pressure due to degenerate electron gas, and thereby the WDs form a stable equilibrium. \cite{1931ApJ....74...81C,1935MNRAS..95..207C} first proposed the idea of the existence of a mass-limit of WDs. He showed that for a carbon-oxygen non-rotating non-magnetized WD, the maximum possible mass is $\sim1.4M_\odot$, popularly known as the Chandrasekhar mass-limit. The theory of general relativity and basic quantum mechanics are sufficient to explain this mass-limit, though Newton's law is enough to understand its existence. Beyond this mass-limit, the pressure balance is no longer sustained, and a WD blows up to produce a type Ia supernova (SNIa). The luminosities of SNeIa are very important as they are used as one of the standard candles to measure cosmological distances. However, during the past couple of decades, the inference of super-Chandrasekhar WDs has been considered as one of the revolutionary discoveries in astrophysics. \cite{2006Natur.443..308H} first reported an over-luminous SNIa, named SN 2003fg, with the content of Nickel mass itself is $\sim 1.3M_\odot$, and thereby they predicted that progenitor mass of the WD for that SNIa is $\sim 2.1M_\odot$. Eventually, several similar over-luminous SNeIa have been discovered, which imply that the progenitor mass of WDs could be as high as $\sim2.8M_\odot$ \citep{2007ApJ...669L..17H,2009ApJ...707L.118Y,2010ApJ...715.1338Y,2010ApJ...714.1209T,2010ApJ...713.1073S,2011MNRAS.410..585S,2011MNRAS.412.2735T,2012ApJ...757...12S}. These WDs are eventually termed as super-Chandrasekhar WD, as they violate the Chandrasekhar mass-limit significantly. This violation of the Chandrasekhar mass-limit challenges use of the standard candle from the luminosities of SNeIa.

While the existence of such a massive WD progenitor for SNeIa was attempted to argue by the double degenerate scenario, numerical simulations of massive WD merger never could lead to the observationally inferred progenitor mass as high as $2.8M_\odot$. Such double degenerate evolutions always produced the off-center ignition and formation of a neutron star rather than a (over-luminous) SNIa (e.g. \citealt{2004ApJ...615..444S,2006MNRAS.373..263M}). Although there are limitations in numerical simulations including chosen mass of component WDs, recently \cite{2019MNRAS.483..263W} showed that the final outcome of WD mergers practically is not influenced by initial WD masses, it primarily depends on the mass-accretion rates during mergers. In a single degenerate scenario of accreting differentially rotating WDs in close binaries of a normal companion, \cite{2009ApJ...702..686C} showed that very massive ($> 1.7M_\odot$) progenitor is not possible to be formed. Hence, all the conventional pictures have yet failed to explain the existence of super-Chandrasekhar progenitor WDs.

\cite{2012MPLA...2750084K} first showed that in the presence of a high 
magnetic field, which forms Landau levels (microscopic effect) in the plane 
perpendicular to the magnetic field axis, super-Chandrasekhar WDs are 
possible; and it leads to a new mass-limit $\sim2.6M_\odot$ 
(\citealt{2013PhRvL.110g1102D}, and the references therein). 
Further, Mukhopadhyay and his collaborators also showed that the macroscopic 
effect of the magnetic field (e.g., magnetic field pressure, magnetic field 
geometry) can also increase the mass of WDs significantly 
\citep{2015MNRAS.454..752S,2019MNRAS.490.2692K}. This idea was verified by, 
e.g., \cite{2015PhRvD..92h3006F,2015RAA....15.1735M,2016MNRAS.456.3375B}, to
name a few.
Similarly, many other researchers proposed different theories, such as 
modified gravity \citep{2018JCAP...09..007K,2017EPJC...77..871C}, generalized 
Heisenberg uncertainty principle \citep{2018JCAP...09..015O}, charged WDs 
\citep{2014PhRvD..89j4043L}, non-commutative geometry 
\citep{2019arXiv191200900K}, to mention a very few, to explain the 
super-Chandrasekhar WDs. Each of these theories gives rise to different 
mass-radius relations for the WDs. However, since none of such 
super-Chandrasekhar WDs have so far been detected directly, the astroseismology 
of such WDs cannot be carried out. Hence, it has not yet been 
possible to single out which one of those theories is the theory behind
the super-Chandrasekhar WDs. It has already been argued that if one considers 
the idea of magnetized super-Chandrasekhar WDs, such WDs possess very less 
thermal luminosity \citep{2018MNRAS.477.2705B}, and hence they have not been 
detected so far by any of the surveys, such as GAIA, Kepler, SDSS. The 
maximum observed magnetic field in an isolated WD is $\sim 10^9$ G 
\citep{2000MNRAS.317..310H,2015SSRv..191..111F}. We have argued that if the 
magnetized WDs have a misalignment between the rotation and magnetic axes 
(same as the configuration of a pulsar), apart from dipole radiation, it can 
emit significant amount of gravitational radiation, which might be detected by 
the future space-based gravitational wave (GW) detectors, such as LISA, 
DECIGO, BBO. Thereby it would be a direct detection of 
super-Chandrasekhar WDs \citep{2019MNRAS.490.2692K}. In this paper, we address 
the timescales related to the dipole and gravitational radiations for these 
pulsating WDs.

Unlike WDs, calculating various timescales for neutron star (NS) pulsars is not a new problem. Pulsars are generally rotating magnetized NSs with the magnetic and rotation axes not aligned with each other. Radio astronomers estimate the lifetime of a pulsar just by calculating its observed period ($P$) and the rate of change of period ($\dot{P}$). The characteristic age of a pulsar is therefore given by $P/2\dot{P}$ \citep{2008LRR....11....8L}. However, this formula is valid, if one considers that the angle between magnetic and rotation axes of a pulsar does not vary throughout its lifetime. However, in practice, due to emission of radiation, this angle is expected to change. \cite{1970ApL.....5...21M} and \cite{1970ApJ...159L..81D} calculated the pulsar timescales simultaneously considering the variations of the angle as well as the spin period of the pulsar emitting dipole radiation, based on the torques calculated earlier by \cite{1955AnAp...18....1D}. Eventually, various researchers used this formalism to solve different properties of pulsars, such as, braking index \citep{1970ApJ...160L..11G,1980ApJ...236..245F,1981Natur.292..811H,1985ApJ...299..706G}, evolution of pulsar magnetic field \citep{1977ApJ...215..302F,1981A&A....98..207K}.
\cite{1970ApJ...161L.137C} included the quadrupolar radiation along-with the dipole radiation and recalculated the various aspects of NSs. All these calculations assumed spherical stars, which, however, are not true in the presence of magnetic field and rotation. \cite{2000MNRAS.313..217M} generalized the equations and applied them for non-spherical NSs. Similarly, these formulae have again been modified considering plasma filled magnetosphere rather than vacuum magnetosphere \citep{2006ApJ...648L..51S,2014MNRAS.441.1879P}. More recently, this formalism has been used to describe the highly magnetized NSs known as magnetars \citep{2018MNRAS.480.4402L,2019ApJ...886....5S,2019arXiv191014336L}.

As mentioned earlier, in this paper, we investigate the time for which a WD 
pulsar can emit dipole and gravitational (quadrupole) radiations, i.e., the 
timescale after which either the magnetic and rotation axes align with each 
other, or the WD stops rotating. This exploration is essential because we 
argued in the earlier paper (\citealt{2019MNRAS.490.2692K}) that the future space-based GW detectors can 
detect the pulsating super-Chandrasekhar WDs. It raises an immediate question 
on the timescale over which we can observe such massive WDs, and in this 
paper, we investigate such timescales for the first time in the case of WD 
pulsars. The plan of the paper is as follows. In Section \ref{Model of 
pulsating white dwarf}, we discuss the fundamental physics behind dipole and 
quadrupole luminosities and thereby formulate the problem. In Section 
\ref{Timescales of magnetized WDs}, we discuss the timescales of various 
possible types of pulsating WDs (regular as well as super-Chandrasekhar) based 
on our model and certain basic properties of GW emitted by the 
isolated magnetized WDs, before we conclude in Section \ref{Conclusions}.


\section{Model of pulsating white dwarf}\label{Model of pulsating white dwarf}

\begin{figure}
\centering
\includegraphics[scale=0.5]{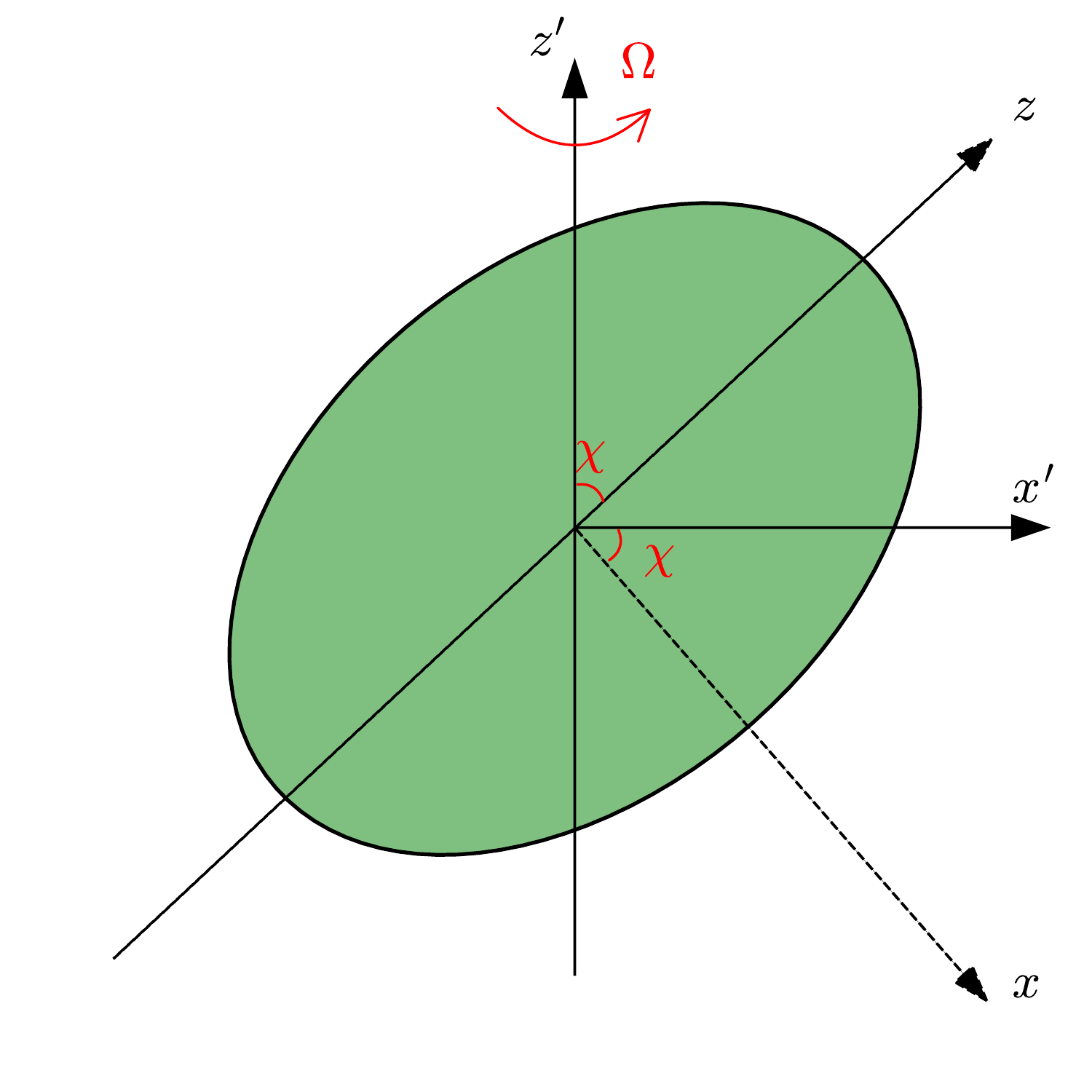}
\caption{Cartoon diagram of a pulsar with $z'$ being the rotational axis and $z$ the magnetic field axis.}
\label{Ellipse}
\end{figure}

Since this paper is based on WD pulsars, we, hereafter, mostly concentrate on 
properties of WDs rather than NSs. It is, of course, well known that the 
number of detected WD pulsars is very less as compared to that of NS pulsars. 
Some of well known WD pulsars are, e.g., AE Aquarii \citep{1987ApJ...323L.131B}, AR 
Scorpii \citep{2016Natur.537..374M}. Figure \ref{Ellipse} shows a cartoon 
diagram of a pulsar with $z'$ being the rotational axis and $z$ the magnetic 
field axis, where the angle between these two axes is $\chi$. It is also known 
for a long time that the magnetic field, as well as rotation, deforms the 
shape and size of any stars \citep{2002PhRvD..66h4025C,2004ApJ...600..296I,
2008PhRvD..78d4045K,2012MNRAS.427.3406F,2015MNRAS.447.3475M,2016MNRAS.459.3407S}. 
Toroidal magnetic field makes a star prolate along with enlarging its size, 
whereas poloidal magnetic field deforms a star to an oblate shape as well as 
reduces its size. Rotation also has similar effects as for poloidal field, 
except that it increases the equatorial radius of the star due to centrifugal 
force \citep{2002PhRvD..66h4025C,2004ApJ...600..296I,2008PhRvD..78d4045K,
2012MNRAS.427.3406F,2015MNRAS.447.3475M,2016MNRAS.459.3407S,2015MNRAS.454..752S,
2019MNRAS.490.2692K}. Hence, the simultaneous presence of magnetic field as 
well as rotation, provided there is a misalignment between their respective 
axes, makes the WD a tri-axial system, which can efficiently produce dipole as 
well as gravitational radiations. It has already been argued that the 
space-based GW detectors, such as LISA, DECIGO, BBO, can detect the gravitational radiation produced by 
such an isolated magnetized WD pulsar in the future 
\citep{2019MNRAS.490.2692K,2020MNRAS.492.5949S}. The dimensionless 
amplitudes of the two polarizations of the GW at a time $t$ are given by 
\citep{1996A&A...312..675B, 1979PhRvD..20..351Z}
\begin{equation}\label{gravitational polarization}
\begin{aligned}
h_+ &= h_0\sin\chi\left[\frac{1}{2}\cos i \sin i\cos\chi\cos\Omega t-\frac{1+\cos^2i}{2}\sin\chi\cos2\Omega t\right],\\
h_\times &= h_0\sin\chi\left[\frac{1}{2}\sin i\cos\chi\sin\Omega t-\cos i\sin\chi\sin2\Omega t\right],
\end{aligned}
\end{equation}
with 
\begin{equation}\label{grav_wave_amplitude}
h_0 = \frac{4G}{c^4}\frac{\Omega^2\epsilon I_{xx}}{d},
\end{equation}
where $G$ is Newton's gravitational constant, $c$ is the speed of light, $\Omega$ is the angular frequency, $d$ is the distance between the detector and the source, $i$ is the angle between the rotation axis of the object and our line of sight, and $\epsilon = |I_{zz}-I_{xx}|/I_{xx}$ with $I_{xx}$ and $I_{zz}$ being the moments of inertia of the WD about $x-$ and $z-$ axes respectively. It is evident from equations \eqref{gravitational polarization} that the amplitude of GW detected by the detector is always one to two orders of magnitude less than $h_0$ depending on the value of $\chi$ and $i$.

Since a pulsating WD can emit both dipole and gravitational radiations simultaneously, it is associated with both the dipole and quadrupolar luminosities. The dipole luminosity for an axisymmetric WD is given by \citep{2000MNRAS.313..217M}
\begin{equation}
L_\text{D} = \frac{B_p^2 R_p^6 \Omega^4}{2c^3} \sin^2\chi ~F(x_0),
\end{equation}
where $x_0=R_0 \Omega/c$, $B_p$ is the strength of the magnetic field at the pole, $R_p$ is the radius of the pole and $R_0$ is the average radius of the WD. The function $F(x_0)$ is defined as
\begin{equation}
F(x_0) = \frac{x_0^4}{5\left(x_0^6 - 3x_0^4 + 36\right)} + \frac{1}{3\left(x_0^2 + 1\right)}.
\end{equation}
Similarly, the quadrupolar GW luminosity is given by \citep{1979PhRvD..20..351Z}
\begin{equation}
L_\text{GW} = \frac{2G}{5c^5} (I_{zz}-I_{xx})^2 \Omega^6 \sin^2\chi \left(1+15\sin^2\chi\right).
\end{equation}
It is important to note that this formula is valid if $\chi$ is very small. The total luminosity of a WD is due to both dipole and gravitational radiations. Hence the changes in $\Omega$ and $\chi$ with respect to time are dependent both in $L_\text{D}$ and $L_\text{GW}$. The variations of $\Omega$ and $\chi$ with respect to time are given by \citep{1970ApJ...161L.137C,2000MNRAS.313..217M}
\begin{equation}\label{Eq: radiation1}
\begin{split}
\frac{d(\Omega I_{z'z'})}{dt} &= -\frac{2G}{5c^5} \left(I_{zz}-I_{xx}\right)^2 \Omega^5 \sin^2\chi \left(1+15\sin^2\chi\right)\\ &- \frac{B_p^2 R_p^6 \Omega^3}{2c^3}\sin^2\chi ~F(x_0),\\
\end{split}
\end{equation}
\begin{equation}\label{Eq: radiation2}
\begin{split}
I_{z'z'} \frac{d\chi}{dt} &= -\frac{12G}{5c^5} \left(I_{zz}-I_{xx}\right)^2 \Omega^4 \sin^3\chi \cos\chi \\ &- \frac{B_p^2 R_p^6 \Omega^2}{2c^3}\sin\chi \cos\chi ~F(x_0),
\end{split}
\end{equation}
where $I_{z'z'}$ is the moment of inertia of the body about $z'-$ axis. Considering small angle approximation, it can be expanded as
\begin{equation}
I_{z'z'} = I_{zz} \cos^2\chi + I_{xx} \sin^2\chi.
\end{equation}
The set of equations \eqref{Eq: radiation1} and \eqref{Eq: radiation2} needs to be solved simultaneously to obtain the timescale over which a WD can radiate.

To solve the equations \eqref{Eq: radiation1} and \eqref{Eq: radiation2}, one 
needs to supply the various quantities, such as $I_{xx}$, $I_{zz}$, $B_p$, 
$R_p$ at the initial time. We use a numerical code named 
{\it XNS},
developed to study the structure of NSs primarily \citep{2014MNRAS.439.3541P}, 
which, however, was appropriately modified for WDs \citep{2015MNRAS.454..752S}. 
This code provides the axisymmetric equilibrium (not necessarily stable 
equilibrium) structure of a stellar body. The advantage of this code is that 
it can give equilibrium solution of uniformly as well as differentially 
rotating WDs in the presence of toroidal or poloidal or twisted-torus magnetic 
fields. However, one needs to supply the equation of state (EoS) in the 
polytropic form, i.e., $\mathcal{P}=K\rho^\Gamma$ with $\mathcal{P}$ being the pressure and $\rho$ 
being the density. In case of WDs with central density $\rho_c$ high and 
magnetic field $\lesssim 10^{15}$ G, the EoS becomes relativistic, which 
implies that $\Gamma \approx 4/3$ and $K \approx (1/8)(3/\pi)^{1/3}hc/(\mu_e 
m_H)^{4/3}$, where $h$ is Planck's constant, $\mu_e$ is the mean molecular 
weight per electron, and $m_H$ is the mass of the hydrogen atom. Moreover, 
{\it XNS} assumes that the rotation and magnetic field axes are in the same 
direction, i.e., $\chi=0$, due to its axisymmetric nature of the algorithm 
mentioned above. However, in case of pulsars, since $\chi \neq 0$ is a 
necessary condition, we assume $\chi$ to be small so that the values of all 
the calculated quantities using the {\it XNS} code, such as mass, radius, 
moment of inertia, are almost valid even if $\chi \neq 0$ 
(\citealt{2019MNRAS.490.2692K}).


\section{Timescales of magnetized white dwarfs}\label{Timescales of magnetized WDs}

\begin{figure}
\gridline{\fig{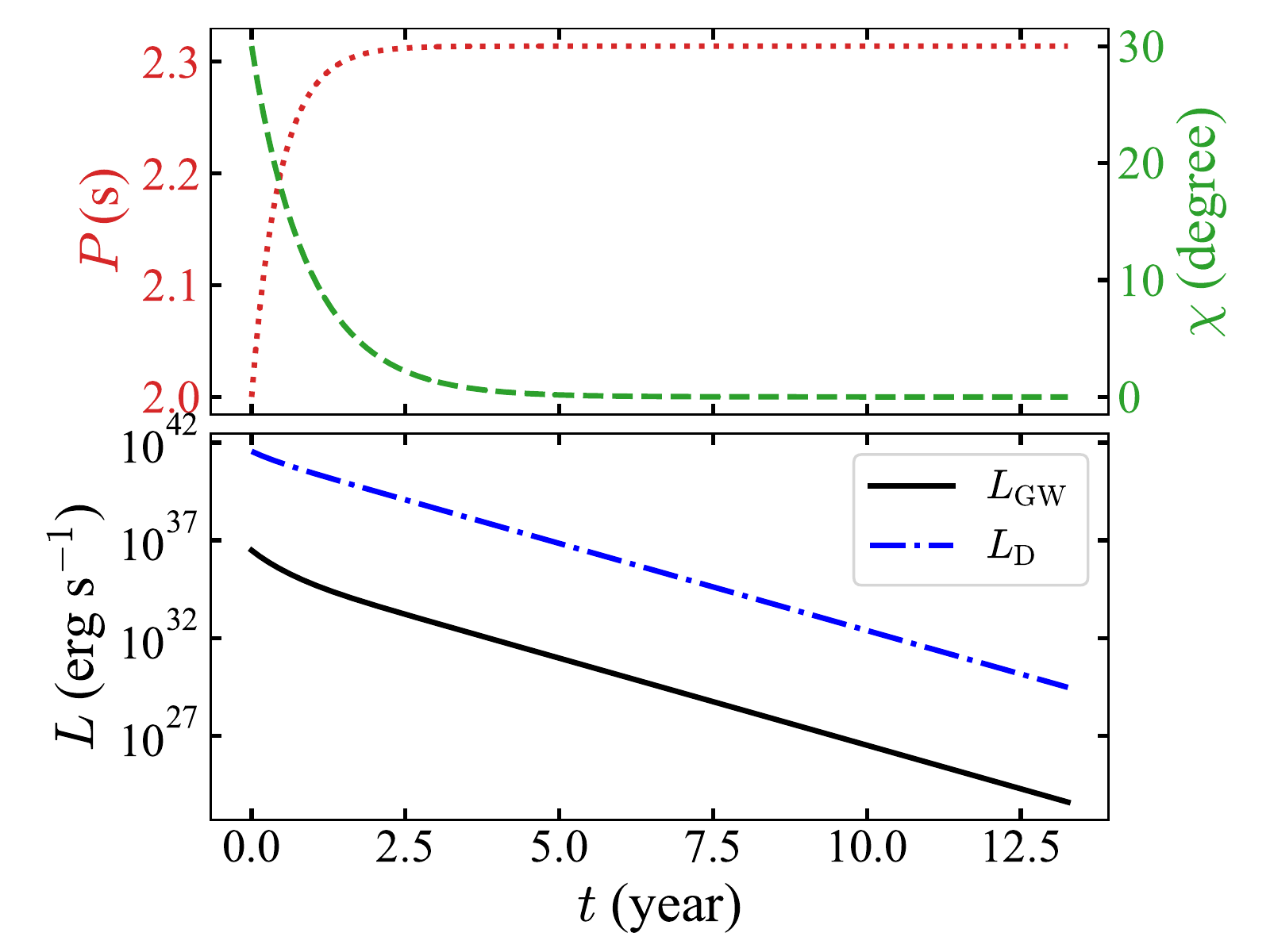}{0.5\textwidth}{(a) $L_{\text{D}}$ is higher than $L_{\text{GW}}$.}}
\gridline{\fig{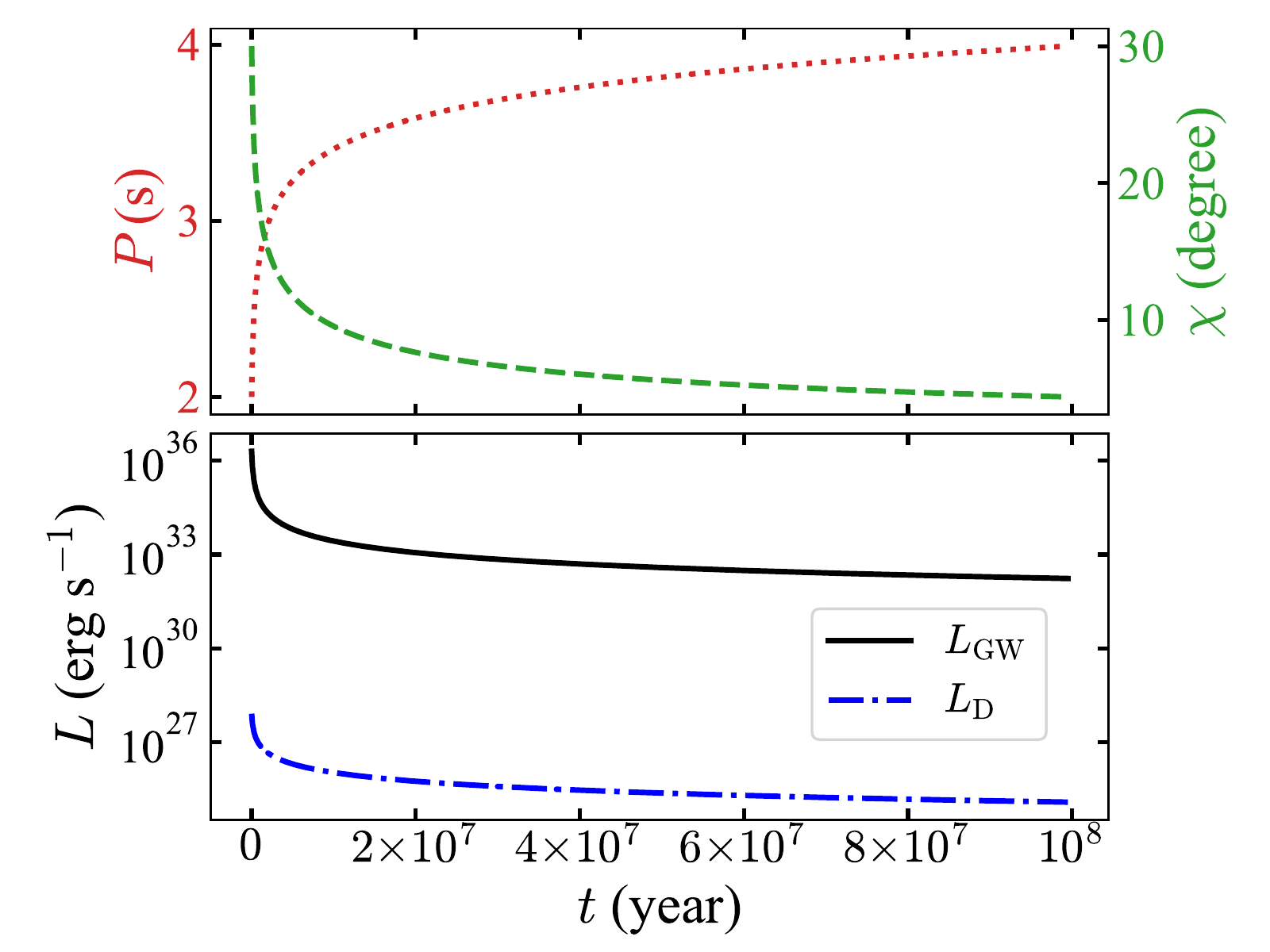}{0.5\textwidth}{(b) $L_{\text{GW}}$ is higher than $L_{\text{D}}$.}}
\caption{Variations of $L_{\text{D}}$, $L_{\text{GW}}$, $P$ and $\chi$ with respect to time. Dotted red and dashed green lines show the variations of $P$ and $\chi$ respectively.}
\label{Fig: Luminosity}
\end{figure}

Since massive WDs can only be formed when $\rho_c$ is high, we choose $\rho_c$ 
to be $10^9$, $10^{10}$ and $2\times 10^{10}$ g cm$^{-3}$ for our 
calculations, where the relativistic EoS mentioned above, is perfectly valid. 
For each of these $\rho_c$, we choose various combinations of $\Omega$ and 
$B_p$ along with the initial angle $\chi$ to be $30\degree$ so that we have an 
idea about the timescales for all possible types of WDs behaving as 
pulsars\footnote{This choice of $\chi = 30\degree$ may still be high, when the 
model equations are valid for small $\chi$ limit. But as our aim here is to 
explore the timescale to decay $\chi$ from its initial value, this choice is 
made and seems to be acceptable. Even if we choose a smaller value of the initial $\chi$, say $5\degree$, the timescale will alter slightly.}. Moreover, we first choose, in the following 
exploration, WDs possessing only purely poloidal magnetic field so that we can 
treat them as oscillating dipoles, and the formula for dipole luminosity is 
valid. Subsequently, we also choose a case with toroidal magnetic fields 
appropriately. While calculating the timescales, we define $t_{10}$, which is 
the time required for a WD to reach 10 orders of less luminosity than what it 
originally possessed at its birth.

\begin{table*}
\caption{Poloidal magnetic field with $\rho_c = 2 \times 10^{10}$ g cm$^{-3}$. $t_{10}$ is the time to decay $L_\text{initial}$ to $10^{-10}L_\text{initial}$. $t_{vl}$ means the timescale is very large.}
\label{Table: Poloidal 2_10}
\centering
\begin{tabular}{|l|l|l|l|l|l|l|l|l|l|}
\hline
\shortstack{$M$ \\ ($M_\odot$)} & \shortstack{$R_P$ \\ (km)} & $B_p$ (G) & $P$ (s) & ME/GE & KE/GE & $L_\text{GW}$ (erg/s) & $L_\text{D}$ (erg/s) & $h_0$ & $t_{10}$ (year)\\
\hline\hline
1.42 & 1200.5 & $8.9 \times 10^{11}$ & 2.0 & $6.1 \times 10^{-4}$ & $2.5 \times 10^{-3}$ & $3.1 \times 10^{36}$ & $3.6 \times 10^{41}$ & $5.2\times 10^{-22}$ & $1.1 \times 10^{1}$ \\
1.42 & 1209.4 & $1.4 \times 10^{9}$ & 2.0 & $1.4 \times 10^{-9}$ & $2.5 \times 10^{-3}$ & $2.0 \times 10^{36}$ & $8.1 \times 10^{35}$ & $4.2\times 10^{-22}$ & $5.4 \times 10^{6}$ \\
1.42 & 1209.4 & $1.3 \times 10^{5}$ & 2.0 & $1.4 \times 10^{-17}$ & $2.5 \times 10^{-3}$ & $2.0 \times 10^{36}$ & $8.1 \times 10^{27}$ & $4.2\times 10^{-22}$ & $t_{vl}$ \\
1.41 & 1218.2 & $8.4 \times 10^{11}$ & 10.0 & $6.0 \times 10^{-4}$ & $9.6 \times 10^{-5}$ & $9.1 \times 10^{30}$ & $5.5 \times 10^{38}$ & $4.5\times 10^{-24}$ & $2.8 \times 10^{2}$ \\
1.41 & 1218.2 & $1.3 \times 10^{9}$ & 10.0 & $1.4 \times 10^{-9}$ & $9.6 \times 10^{-5}$ & $1.4 \times 10^{29}$ & $1.2 \times 10^{33}$ & $5.7\times 10^{-25}$ & $1.2 \times 10^{8}$ \\
1.41 & 1218.2 & $1.3 \times 10^{5}$ & 10.0 & $1.4 \times 10^{-17}$ & $9.6 \times 10^{-5}$ & $1.4 \times 10^{29}$ & $1.2 \times 10^{25}$ & $5.7\times 10^{-25}$ & $t_{vl}$ \\
1.41 & 1218.2 & $8.4 \times 10^{11}$ & 100.0 & $6.0 \times 10^{-4}$ & $9.6 \times 10^{-7}$ & $6.7 \times 10^{24}$ & $5.5 \times 10^{34}$ & $3.8\times 10^{-26}$ & $2.8 \times 10^{4}$ \\
1.41 & 1218.2 & $1.3 \times 10^{9}$ & 100.0 & $1.4 \times 10^{-9}$ & $9.6 \times 10^{-7}$ & $2.5 \times 10^{21}$ & $1.2 \times 10^{29}$ & $7.4\times 10^{-28}$ & $1.2 \times 10^{10}$ \\
1.41 & 1218.2 & $1.3 \times 10^{5}$ & 100.0 & $1.4 \times 10^{-17}$ & $9.6 \times 10^{-7}$ & $2.5 \times 10^{21}$ & $1.2 \times 10^{21}$ & $7.4\times 10^{-28}$ & $1.3 \times 10^{18}$ \\
\hline
\end{tabular}
\end{table*}

\begin{table*}
\caption{Poloidal magnetic field with $\rho_c = 10^{10}$ g cm$^{-3}$. $t_{10}$ is the time to decay $L_\text{initial}$ to $10^{-10}L_\text{initial}$. $t_{vl}$ means the timescale is very large.}
\label{Table: Poloidal 1_10}
\centering
\begin{tabular}{|l|l|l|l|l|l|l|l|l|l|}
\hline
\shortstack{$M$ \\ ($M_\odot$)} & \shortstack{$R_P$ \\ (km)} & $B_p$ (G) & $P$ (s) & ME/GE & KE/GE & $L_\text{GW}$ (erg/s) & $L_\text{D}$ (erg/s) & $h_0$ & $t_{10}$ (year)\\
\hline\hline
1.44 & 1510.6 & $7.4 \times 10^{11}$ & 2.0 & $1.0 \times 10^{-3}$ & $5.1 \times 10^{-3}$ & $3.4 \times 10^{37}$ & $9.8 \times 10^{41}$ & $1.7\times 10^{-21}$ & $6.7 \times 10^{0}$ \\
1.44 & 1519.5 & $1.1 \times 10^{9}$ & 2.0 & $2.2 \times 10^{-9}$ & $5.1 \times 10^{-3}$ & $2.4 \times 10^{37}$ & $2.2 \times 10^{36}$ & $1.5\times 10^{-21}$ & $3.8 \times 10^{6}$ \\
1.44 & 1519.5 & $1.1 \times 10^{5}$ & 2.0 & $2.2 \times 10^{-17}$ & $5.1 \times 10^{-3}$ & $2.4 \times 10^{37}$ & $2.2 \times 10^{28}$ & $1.5\times 10^{-21}$ & $t_{vl}$ \\
1.42 & 1537.2 & $6.8 \times 10^{11}$ & 10.0 & $9.7 \times 10^{-4}$ & $2.0 \times 10^{-4}$ & $6.7 \times 10^{31}$ & $1.5 \times 10^{39}$ & $1.2\times 10^{-23}$ & $1.9 \times 10^{2}$ \\
1.42 & 1546.1 & $1.0 \times 10^{9}$ & 10.0 & $2.2 \times 10^{-9}$ & $2.0 \times 10^{-4}$ & $1.8 \times 10^{30}$ & $3.3 \times 10^{33}$ & $2.0\times 10^{-24}$ & $7.5 \times 10^{7}$ \\
1.42 & 1546.1 & $1.0 \times 10^{5}$ & 10.0 & $2.2 \times 10^{-17}$ & $2.0 \times 10^{-4}$ & $1.8 \times 10^{30}$ & $3.3 \times 10^{25}$ & $2.0\times 10^{-24}$ & $t_{vl}$ \\
1.42 & 1537.1 & $6.7 \times 10^{11}$ & 100.0 & $9.7 \times 10^{-4}$ & $2.0 \times 10^{-6}$ & $4.6 \times 10^{25}$ & $1.5 \times 10^{35}$ & $1.0\times 10^{-25}$ & $1.7 \times 10^{4}$ \\
1.42 & 1546.1 & $1.0 \times 10^{9}$ & 100.0 & $2.2 \times 10^{-9}$ & $2.0 \times 10^{-6}$ & $5.4 \times 10^{21}$ & $3.3 \times 10^{29}$ & $1.1\times 10^{-27}$ & $7.5 \times 10^{9}$ \\
1.42 & 1546.1 & $1.0 \times 10^{5}$ & 100.0 & $2.2 \times 10^{-17}$ & $2.0 \times 10^{-6}$ & $5.4 \times 10^{21}$ & $3.3 \times 10^{21}$ & $1.1\times 10^{-27}$ & $8.1 \times 10^{17}$ \\
\hline
\end{tabular}
\end{table*}

\begin{table*}
\caption{Poloidal magnetic field with $\rho_c = 10^9$ g cm$^{-3}$. $t_{10}$ is the time to decay $L_\text{initial}$ to $10^{-10}L_\text{initial}$. $t_{vl}$ means the timescale is very large.}
\label{Table: Poloidal 1_9}
\centering
\begin{tabular}{|l|l|l|l|l|l|l|l|l|l|}
\hline
\shortstack{$M$ \\ ($M_\odot$)} & \shortstack{$R_P$ \\ (km)} & $B_p$ (G) & $P$ (s) & ME/GE & KE/GE & $L_\text{GW}$ (erg/s) & $L_\text{D}$ (erg/s) & $h_0$ & $t_{10}$ (year) \\
\hline\hline
1.48 & 3021.3 & $6.6 \times 10^{11}$ & 5.3 & $1.1 \times 10^{-2}$ & $7.4 \times 10^{-3}$ & $1.8 \times 10^{37}$ & $1.0 \times 10^{42}$ & $3.3\times 10^{-21}$ & $4.6 \times 10^{0}$ \\
1.46 & 3216.2 & $1.2 \times 10^{9}$ & 5.3 & $5.1 \times 10^{-8}$ & $7.4 \times 10^{-3}$ & $3.6 \times 10^{36}$ & $5.0 \times 10^{36}$ & $1.5\times 10^{-21}$ & $9.4 \times 10^{5}$ \\
1.46 & 3216.2 & $1.2 \times 10^{5}$ & 5.3 & $5.1 \times 10^{-16}$ & $7.4 \times 10^{-3}$ & $3.6 \times 10^{36}$ & $5.0 \times 10^{28}$ & $1.5\times 10^{-21}$ & $t_{vl}$ \\
1.46 & 3092.1 & $5.9 \times 10^{11}$ & 10.0 & $1.1 \times 10^{-2}$ & $1.9 \times 10^{-3}$ & $1.6 \times 10^{35}$ & $7.3 \times 10^{40}$ & $6.0\times 10^{-22}$ & $1.7 \times 10^{1}$ \\
1.43 & 3287.0 & $1.1 \times 10^{9}$ & 10.0 & $4.9 \times 10^{-8}$ & $1.9 \times 10^{-3}$ & $4.5 \times 10^{33}$ & $3.5 \times 10^{35}$ & $1.0\times 10^{-22}$ & $3.3 \times 10^{6}$ \\
1.43 & 3287.0 & $1.1 \times 10^{5}$ & 10.0 & $4.9 \times 10^{-16}$ & $1.9 \times 10^{-3}$ & $4.5 \times 10^{33}$ & $3.5 \times 10^{27}$ & $1.0\times 10^{-22}$ & $t_{vl}$ \\
1.45 & 3109.8 & $5.7 \times 10^{11}$ & 100.0 & $1.1 \times 10^{-2}$ & $1.9 \times 10^{-5}$ & $1.1 \times 10^{29}$ & $7.1 \times 10^{36}$ & $5.0\times 10^{-2 4}$ & $1.7 \times 10^{3}$ \\
1.43 & 3304.8 & $1.0 \times 10^{9}$ & 100.0 & $4.8 \times 10^{-8}$ & $1.9 \times 10^{-5}$ & $5.8 \times 10^{22}$ & $3.4 \times 10^{31}$ & $3.6\times 10^{-27}$ & $3.4 \times 10^{8}$ \\
1.43 & 3304.8 & $1.0 \times 10^{5}$ & 100.0 & $4.8 \times 10^{-16}$ & $1.9 \times 10^{-5}$ & $5.7 \times 10^{22}$ & $3.4 \times 10^{23}$ & $3.6\times 10^{-27}$ & $3.4 \times 10^{16}$ \\
\hline
\end{tabular}
\end{table*}

All the different combinations of $\Omega$ and $B_p$ for different $\rho_c$ 
are given in Tables \ref{Table: Poloidal 2_10}, \ref{Table: Poloidal 1_10} and 
\ref{Table: Poloidal 1_9} with the respective mass $M$, $R_p$ and $h_0$. We 
assume throughout the distance of the source from the detector $d = 100$ pc. 
Here we primarily restrict the magnetic to gravitational energies ratio (ME/GE) as 
well as the kinetic to gravitational energies ratio (KE/GE) to less than $\sim 10^{-2}$ 
so that the magnetized WDs are surely stable \citep{1989MNRAS.237..355K,2009MNRAS.397..763B}. 
It is important to note that with these values of ME/GE and KE/GE, 
a WD cannot possess mass significantly more than the Chandrasekhar 
mass-limit. However, this limit may be relaxed in a suitable mixed 
field configuration leading to super-Chandrasekhar WDs, which is 
beyond the scope of the present work as {\it XNS} cannot 
handle a rotating star with a suitable and/or an equal fractions of mixed field configuration. 
It is of course long known that the stars containing purely toroidal 
or purely poloidal magnetic fields are unstable (\citealt{tayler1,tayler2}). However, in the 
present work, our aim is not to study the stability analysis, 
and the {\it code} we relied upon cannot handle a rotating star with suitable mixed field configurations. Hence, 
purely poloidal or purely toroidal magnetic fields, maintaining ME/GE 
limit mentioned above, are valid approximations of poloidally dominated or toroidally 
dominated mixed field configurations.
Below we discuss the time evolutions of rotational period, angle between magnetic
and rotational axes and various luminosities of WDs. 

\subsection{Case I: $L_\text{D}>>L_\text{GW}$} \label{LD>>LGW}

Since $L_\text{D}$ increases with an increase in the magnetic field, it is 
understood from the tables that the WDs possessing high value of the magnetic 
field have $L_\text{D}>>L_\text{GW}$. Since the luminosity is dominated by 
$L_\text{D}$ and $L\propto dE/dt$, the timescale is governed by $L_\text{D}$. 
Moreover, the total luminosity of a WD decreases with time either due to 
decrease in $\chi$ or decrease in $\Omega$. Whenever $L_\text{D}>>L_\text{GW}$, 
$\chi$ decreases much faster as compared to $\Omega$. For $L_\text{D}>>L_\text{GW}$, 
the equations \eqref{Eq: radiation1} and \eqref{Eq: radiation2} can be 
approximated as follows
\begin{align}\label{Eq: radiation_D1}
I_{z'z'}\frac{d\Omega}{dt} &= - \frac{B_p^2 R_p^6 \Omega^3}{2c^3}\sin^2\chi ~F(x_0),\\
\label{Eq: radiation_D2}
I_{z'z'} \frac{d\chi}{dt} &= - \frac{B_p^2 R_p^6 \Omega^2}{2c^3}\sin\chi \cos\chi ~F(x_0),
\end{align}
assuming $I_{z'z'}$ not changing with time. Let us denote the timescale for the change in $\Omega$ to be $T_\Omega$ and that for $\chi$ to be $T_\chi$. Integrating these two equation, we obtain
\begin{align}
T_\Omega &\sim \left(\frac{2I_{z'z'}c^3}{B_p^2 R_p^6 \Omega^2 F(x_0)}\right)\frac{1}{2\sin^2\chi},\\
T_\chi &\sim \left(\frac{2I_{z'z'}c^3}{B_p^2 R_p^6 \Omega^2 F(x_0)}\right) \ln\cot\chi.
\end{align}
In the range $0\degree \leqslant \chi \leqslant 30\degree$, we always have 
$\ln\cot\chi << 1/2\sin^2\chi$, which implies $T_\chi<<T_\Omega$. This proves 
that $\chi$ quickly becomes 0, and the WD starts rotating with a different 
angular velocity than it originally possesses. 
For example, if $M = 1.42M_\odot$, $B_p = 8.9 \times 10^{11}$ G, 
$R_p = 1200$ km, and at $t=0$, $\Omega = \pi$ rad/s, and $\chi = 30\degree$ such that 
$L_\text{D}>>L_\text{GW}$, then $T_\Omega \sim 3$ years and $T_\chi \sim 0.2$ year.
It can also be verified from Figure \ref{Fig: Luminosity}(a). Moreover, 
combining the equations \eqref{Eq: radiation_D1} and \eqref{Eq: radiation_D2}, 
we obtain the differential equation
\begin{align}
\frac{d\Omega}{d\chi} = \Omega \frac{\sin \chi}{\cos \chi}.
\end{align}
Solving this differential equation using the initial condition $\chi = 30\degree$, we obtain
\begin{equation}
\Omega = \frac{\sqrt{3}}{2 \cos\chi}\Omega_0,
\end{equation}
where $\Omega_0$ is the initial angular velocity of the WD. Using this 
formula, one can verify that the final time period would be $\sim 2.3$ s if 
the initial time period is $2$ s, and it is clearly evident from Figure \ref{Fig: Luminosity}(a).

\subsection{Case II: $L_\text{GW}>>L_\text{D}$} \label{LGW>>LD}

If the magnetic field is lower, but not the angular velocity, the WDs have $L_\text{GW}>>L_\text{D}$. In such a case, luminosity decreases slowly and a WD can radiate for a long period of time. For $L_\text{GW}>>L_\text{D}$, the equations \eqref{Eq: radiation1} and \eqref{Eq: radiation2} can be written as
\begin{align}\label{Eq: radiation_GW1}
I_{z'z'}\frac{d\Omega}{dt} &= -\frac{2G}{5c^5} \left(I_{zz}-I_{xx}\right)^2 \Omega^5 \sin^2\chi ~(1+15\sin^2\chi),\\
\label{Eq: radiation_GW2}
I_{z'z'} \frac{d\chi}{dt} &= -\frac{12G}{5c^5} \left(I_{zz}-I_{xx}\right)^2 \Omega^4 \sin^3\chi \cos\chi.
\end{align}
Integrating these two equations, we obtain the timescales of changes in $\Omega$ and $\chi$, given by
\begin{align}
T'_\Omega &\sim \left(\frac{5I_{z'z'}c^5}{2 G \left(I_{zz}-I_{xx}\right)^2 \Omega^4}\right)\frac{1}{4\sin^2\chi(1+15\sin^2\chi)},\\
T'_\chi &\sim \left(\frac{5I_{z'z'}c^5}{2 G \left(I_{zz}-I_{xx}\right)^2 \Omega^4}\right)\frac{1}{12}\left(\frac{1}{\sin^2\chi}+2\ln\cot\chi\right).
\end{align}
In the range $0\degree \leqslant \chi \leqslant 30\degree$, $T'_\Omega \sim T'_\chi$. 
Hence the $\Omega$ and $\chi$ keep varying simultaneously for a long time 
before approaching to zero, which also can be verified from Figure \ref{Fig: Luminosity}(b). 
For instance, if $M = 1.42M_\odot$, $B_p = 1.3 \times 10^5$ G, $R_p = 1209$ km, and at $t=0$,
$\Omega = \pi$ rad/s, and $\chi = 30\degree$ such that $L_\text{GW}>>L_\text{D}$, 
then $T_\Omega \sim 4 \times 10^{12}$ years and $T_\chi \sim 7 \times 10^{12}$ years.
Moreover, combining the equations \eqref{Eq: radiation_GW1} and \eqref{Eq: radiation_GW2}, we obtain
\begin{align}
\frac{d\Omega}{d\chi} = \Omega \frac{1+15\sin^2\chi}{\sin\chi\cos \chi}.
\end{align}
Solving this equation using the initial condition $\chi=30\degree$, we obtain
\begin{equation}
\Omega = \frac{3^8\sin\chi}{2^{15}\cos^{16}\chi}\Omega_0.
\end{equation}
This proves that as $\chi \to 0$, $\Omega \to 0$; unlike the earlier case 
mentioned in \S \ref{LD>>LGW}. Hence the overall timescale is 
determined by the change in both $\Omega$ and $\chi$, and in this case, it 
turns out to be much longer.


\begin{table*}
\caption{Super-Chandrasekhar WDs possessing poloidal magnetic field for $\rho_c = 2 \times 10^{10}$ g cm$^{-3}$. WDs in first three rows follow Chandrasekhar EoS and the rests follow non-commutative EoS. $t_{vl}$ means the timescale is very large.}
\label{Table: super_Chandrasekhar}
\centering
\begin{tabular}{|l|l|l|l|l|l|l|l|l|l|}
\hline
\shortstack{$M$ \\ ($M_\odot$)} & \shortstack{$R_P$ \\ (km)} & $B_p$ (G) & $P$ (s) & ME/GE & KE/GE & $L_\text{GW}$ (erg/s) & $L_\text{D}$ (erg/s) & $h_0$ & $t_{10}$ (year) \\
\hline\hline
1.69 & 748.7 & $3.6 \times 10^{13}$ & 2.0 & $1.0 \times 10^{-1}$ & $2.3 \times 10^{-3}$ & $1.6 \times 10^{39}$ & $3.5 \times 10^{43}$ & $1.2\times 10^{-20}$ & $1.5 \times 10^{-1}$ \\
1.67 & 757.6 & $3.5 \times 10^{13}$ & 10.0 & $1.0 \times 10^{-1}$ & $9.1 \times 10^{-5}$ & $9.4 \times 10^{34}$ & $5.5 \times 10^{40}$ & $4.6\times 10^{-22}$ & $3.7 \times 10^{0}$ \\
1.67 & 757.5 & $3.5 \times 10^{13}$ & 100.0 & $1.0 \times 10^{-1}$ & $9.1 \times 10^{-7}$ & $9.4 \times 10^{28}$ & $5.5 \times 10^{36}$ & $4.6\times 10^{-24}$ & $3.7 \times 10^{2}$ \\
3.18 & 899.3 & $4.9 \times 10^{13}$ & 2.0 & $1.1 \times 10^{-1}$ & $2.3 \times 10^{-3}$ & $1.3 \times 10^{40}$ & $1.9 \times 10^{44}$ & $3.4\times 10^{-20}$ & $7.8 \times 10^{-2}$ \\
3.15 & 917.0 & $4.6 \times 10^{13}$ & 10.0 & $1.1 \times 10^{-1}$ & $9.1 \times 10^{-5}$ & $7.8 \times 10^{35}$ & $3.0 \times 10^{41}$ & $1.3\times 10^{-21}$ & $1.9 \times 10^{0}$ \\
2.65 & 1492.9 & $1.1 \times 10^{8}$ & 2.0 & $6.7 \times 10^{-12}$ & $2.5 \times 10^{-3}$ & $1.6 \times 10^{37}$ & $2.1 \times 10^{34}$ & $1.2\times 10^{-21}$ & $t_{vl}$ \\
2.64 & 1501.8 & $1.1 \times 10^{8}$ & 10.0 & $6.6 \times 10^{-12}$ & $9.6 \times 10^{-5}$ & $1.1 \times 10^{30}$ & $3.2 \times 10^{31}$ & $1.6\times 10^{-24}$ & $1.3 \times 10^{10}$ \\
\hline
\end{tabular}
\end{table*}

\subsection{Super-Chandrasekhar WDs with poloidal magnetic field}\label{super Chandrasekhar Poloidal field}

As we have mentioned above, the chosen values of magnetic field and rotation in 
Tables \ref{Table: Poloidal 2_10}, \ref{Table: Poloidal 1_10} and 
\ref{Table: Poloidal 1_9} cannot give super-Chandrasekhar WDs. However, at the 
time of its birth, a WD may possess a very high magnetic field (may be 
suitable mixed fields), which is even larger than the Schwinger limit of 
$4.414\times10^{13}$ G. This value of the magnetic field can make the WD 
significantly super-Chandrasekhar. If such a WD behaves like a pulsar, it can 
also emit a significant amount of gravitational radiation. However, due to 
high $L_\text{D}$, these WDs cannot emit radiation for a longer 
duration, as $\chi$ 
becomes zero much quickly. It is evident from first three rows of Table 
\ref{Table: super_Chandrasekhar}, where $B_p$ is larger than $10^{13}$ G. Such 
WDs may also be detected by the future GW detectors just for a short duration 
of time (maybe momentarily). Recently, we have proposed the effect of 
non-commutativity on the EoS of the degenerate electrons 
(\citealt{2019PhLB..79734859P}; \citealt{2019arXiv191200900K}). We have shown that if 
non-commutativity is significant in a WD, it can have a mass up to 
$\sim2.6 M_\odot$ even for a static non-magnetized WD. If such WDs also 
possess magnetic field and rotation, they can also emit a significant amount 
of GW, which can also be detected by the upcoming space-based detectors, such 
as LISA, DECIGO, BBO. The last four rows of Table 
\ref{Table: super_Chandrasekhar} show the timescales for the WDs if their EoS 
is governed by non-commutativity. It is evident that even if such WDs possess 
a lower magnetic field, they emit continuous gravitational radiation for a 
fairly longer duration. This will also be a valid test of the presence of 
non-commutativity in WD matter.

\subsection{Effect of WD's birth rate on its detection}\label{WD's birth rate}

It is important to note that the birth rate of WDs can also significantly 
affect the detection of massive WDs. The birth rate of a WD is $\sim 10^{-12}$ 
pc$^{-3}$ year$^{-1}$ \citep{1983Ap&SS..97..305G}, which means within 100 pc 
radius, on average, only one WD is formed in $10^6$ years. Hence if that 
particular WD is super-Chandrasekhar only due to the high magnetic field, 
particularly poloidally dominated (see below \S \ref{super Chandrasekhar Toroidal field}), it 
may not be detected by the detector, or at best, if one is lucky enough, it 
may be detected only for a short duration of time as it loses its spin-down 
luminosity very quickly. However, if non-commutativity prevails as compared to 
the magnetic field, as described in \S \ref{super Chandrasekhar Poloidal field} 
above, such super-Chandrasekhar WDs (with weaker fields
not really affecting the mass) emit a significant amount of 
GW for a more extended period, and the GW detectors can detect them for a 
longer duration. Of course, it is well known that the presence of a purely 
poloidal magnetic field in the WD makes it unstable (\citealt{tayler2}).
Hence, in reality, such WDs should have some toroidal magnetic
field as well.

\begin{table*}
\caption{WDs possessing toroidal magnetic field for $\rho_c = 2 \times 10^{10}$ g cm$^{-3}$. Here $B_\text{max}$ is the strength of the maximum magnetic field in the WD and $R_E$ is the equatorial radius. $t_{vl}$ means the timescale is very large.}
\label{Table: super_Chandrasekhar_toroidal}
\centering
\begin{tabular}{|l|l|l|l|l|l|l|l|l|}
\hline
\shortstack{$M$ \\ ($M_\odot$)} & \shortstack{$R_E$ \\ (km)} & $B_\text{max}$ (G) & $P$ (s) & ME/GE & KE/GE & $L_\text{GW}$ (erg/s) & $h_0$ & $t_{10}$ (year) \\
\hline\hline
1.71 & 2095.4 & $2.6 \times 10^{14}$ & 2.0 & $1.0 \times 10^{-1}$ & $5.4 \times 10^{-3}$ & $3.1 \times 10^{39}$ & $1.7\times 10^{-20}$ & $t_{vl}$ \\
1.44 & 1315.7 & $1.1 \times 10^{14}$ & 2.0 & $1.0 \times 10^{-2}$ & $2.7 \times 10^{-3}$ & $2.6 \times 10^{35}$ & $4.6\times 10^{-22}$ & $t_{vl}$ \\
1.67 & 1767.6 & $2.6 \times 10^{14}$ & 10.0 & $1.0 \times 10^{-1}$ & $2.0 \times 10^{-4}$ & $2.3 \times 10^{36}$ & $7.5\times 10^{-22}$ & $t_{vl}$ \\
1.43 & 1253.7 & $1.1 \times 10^{14}$ & 10.0 & $1.0 \times 10^{-2}$ & $1.0 \times 10^{-4}$ & $6.3 \times 10^{32}$ & $3.7\times 10^{-23}$ & $t_{vl}$ \\
\hline
\end{tabular}
\end{table*}


\subsection{WDs with toroidal magnetic field}\label{super Chandrasekhar Toroidal field}

Whatever calculations we have shown so far above are based on the simplistic 
assumption of purely poloidal field so that we can consistently use the 
formula of $L_\text{D}$. In reality, a WD is stable only if it consists of 
both the toroidal and poloidal components suitably. However, such a suitable 
configuration is not possible to obtain with the help of the {\it XNS} code. 
{\it XNS} can capture a twisted torus configuration which contains significantly
poloidally dominated fields.
Hence we consider a few cases of WDs for $\rho_c = 2\times 10^{10}$ g 
cm$^{-3}$ containing a toroidal magnetic field. Of course, in this case, we 
drop the contributions of the term $L_\text{D}$. In other words, we assume 
that even if the WD possesses any dipole contribution, its effect is much smaller, 
which is similar to the case, as we have mentioned in \S \ref{LGW>>LD} 
with specific estimates. Such a configuration possesses 
super-Chandrasekhar mass, because the toroidal field is dominant at the 
center and, at the surface, it may have a negligible contribution. WDs 
containing mostly toroidal field and a negligible poloidal component is indeed 
a stable configuration. \cite{2014MNRAS.437..675W} showed that such a 
configuration remains stable even after a long time, which satisfies the stability 
criteria given by \cite{2009MNRAS.397..763B}. They also proposed that the 
poloidal field is generated as a by-product of the decay of the toroidal field. 
As in this configuration $L_\text{GW}>>L_\text{D}$, such a magnetized 
super-Chandrasekhar WD can radiate for a long time. 
Table \ref{Table: super_Chandrasekhar_toroidal} shows the timescales for various WDs with a 
purely toroidal magnetic field. In reality, since a WD (even if it is a 
super-Chandrasekhar) possesses both the toroidal and poloidal magnetic fields 
simultaneously, the GW detector may or may not detect it for a longer duration
depending on the strengths of the toroidal and poloidal field components.
For example, if the mixed stable field configuration is toroidally dominated
like what proposed by, e.g., \cite{2014MNRAS.437..675W}, then the GW detector
is able to trace the source for a long duration, as shown by
Figure \ref{Fig: Luminosity}(b). On the other hand, if the field is 
poloidally dominated or even of approximately equal contributions from 
toroidal and poloidal components, then due to the presence of significant
$L_\text{D}$, luminosity decreases very fast with decaying $\chi$ and 
increasing $P$, as shown by Figure \ref{Fig: Luminosity}(a). In this case, it may not be 
detected by a GW detector or it may be detected only for a short duration 
of time. All these plausibilities, however, is not possible to show in this paper due to the 
limitations of the code. Here purely poloidal and purely toroidal magnetic 
field configurations are replicas of poloidally dominated and toroidally 
dominated mixed field configurations respectively.


\subsection{Detectability of isolated magnetized white dwarfs in 
gravitational wave astronomy}

\begin{figure}
\centering
\includegraphics[scale=0.30]{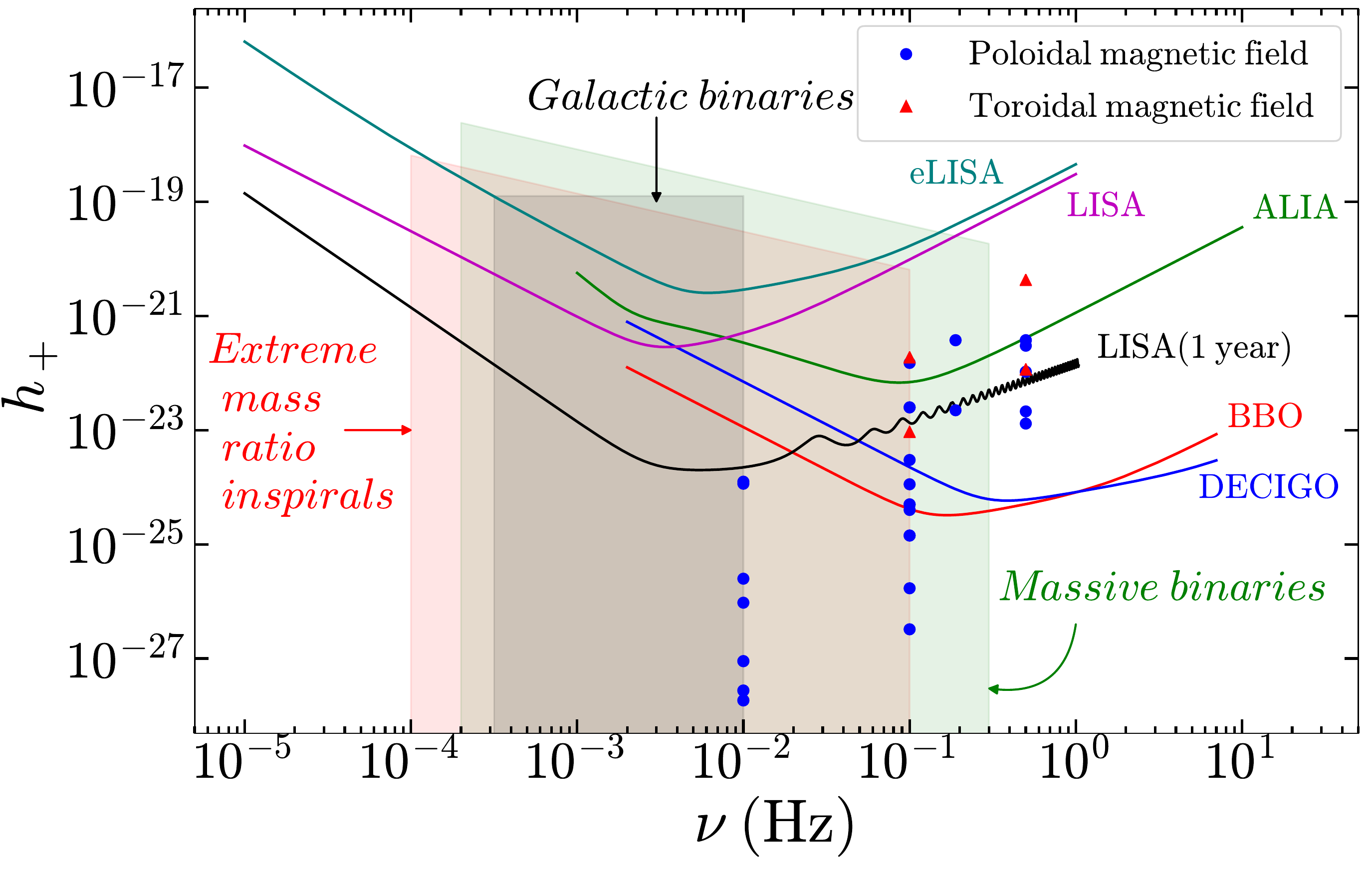}
\caption{Dimensionless GW amplitude for white dwarfs as a function of frequency, as given in Tables \ref{Table: Poloidal 2_10}-\ref{Table: super_Chandrasekhar_toroidal}, along with the sensitivity curves of various detectors. Optimum $i$
is chosen for $\chi$ at $t=0$.}
\label{Fig: Detector_new}
\end{figure}

Let us now briefly discuss the properties of GW strengths emitted 
by the magnetized WDs. A detailed discussion of GW strengths for various 
WDs with different sets of parameters is given by \cite{2019MNRAS.490.2692K}. 
Figure \ref{Fig: Detector_new} shows the dimensionless GW amplitudes 
for the WDs with respect to their frequencies, as given in Tables 
\ref{Table: Poloidal 2_10}-\ref{Table: super_Chandrasekhar_toroidal}, along 
with the sensitivity curves of various detectors\footnote{\url{http://gwplotter.com/} 
and \url{http://www.srl.caltech.edu/~shane/sensitivity/}} 
(\citealt{2009LRR....12....2S,2015CQGra..32a5014M}, 
and the references therein). It is evident that isolated WDs may not be 
detected by LISA directly, which, however, can be detected after integrating 
the signal to noise ratio for 1 year. It is to be noted that being 
larger in size, WDs cannot rotate much faster as of neutron stars, and hence 
ground-based GW detectors, such as LIGO, Virgo, KAGRA, are not
expected to detect the 
isolated WDs. The frequency range of the isolated WDs is different than those
of the stochastic background noise and, hence, nano-hertz GW detectors, 
such as IPTA, SKA, NANOGrav, cannot detect them either. Moreover, from Figure
\ref{Fig: Detector_new} 
it is evident that these isolated WDs are free from the confusion noise of 
the galactic binaries as well as from the extreme mass ratio inspirals (EMRIs). 
However, some of the isolated WDs may fall in the same range of massive 
binaries. Since these sources are different from the isolated WDs, by using 
the specific templates for each of these different sources, the problem of 
confusion noise can be rectified. Moreover, since the luminosity is an 
intrinsic property of the source (which is the amount of energy radiated per 
unit time), the orientation of the source does not matter while calculating 
the timescales.


\section{Conclusions}\label{Conclusions}

In this work, we analyze the timescales related to pulsating WDs. We have 
considered both the dipole and GW luminosities emitted by pulsating WDs, which 
was, in  our knowledge, not explored consistently before this work. We have 
used the {\it XNS} code to model the WDs, primarily containing the poloidal 
magnetic field such that we can treat them as oscillating dipoles. Our target 
has been to calculate the timescale for detecting the super-Chandrasekhar WDs 
through the GW detectors. We have shown that many of such massive WDs have 
higher GW amplitude, and they are well above the signal to noise ratio of the 
GW detectors. If the WDs are massive due to the high poloidal magnetic field, 
i.e., they possess high dipolar luminosities, they cannot be detected for a 
longer duration. However, if a massive WD possesses high toroidal field at the 
center and very less magnetic field at the pole, it can be detected by the 
detectors for a long time. Moreover, if the WDs gain extra mass due to some 
other effects, such as non-commutative geometry, but possess some weaker 
fields, they can also emit 
gravitational radiation continuously for a long time, and the GW detectors 
should easily detect them.

\acknowledgements
S.K. would like to thank Timothy Brandt of the University of California, Santa Barbara, for the useful discussion about the timescale for WDs. B.M. would like to thank Tom Marsh of the University of Warwick for discussion in the conference ``Compact White Dwarf Binaries'', Yerevan, Armenia. B.M. acknowledges a partial support by a project of Department of Science and Technology (DST), India, with Grant No. DSTO/PPH/BMP/1946 (EMR/2017/001226). T.B. is supported by TEAM/2016-3/19 grant from FNP.


\software{{\it XNS} code solves the time independent general relativistic 
magnetohydrodynamic (GRMHD) equations, see \citealt{2014MNRAS.439.3541P}
for development of its latest version, 
url: \url{http://www.arcetri.astro.it/science/ahead/XNS/code.html}}.

\bibliographystyle{yahapj}
\bibliography{mypaper5}

\end{document}